\documentstyle[12pt,epsf]{article}

\def\be{\begin{equation}}
\def\ee{\end{equation}}
\def\bea{\begin{eqnarray}}
\def\eea{\end{eqnarray}}

\newcommand{\Section}[1]{\section{#1}\setcounter{equation}{0}}

\begin{document}

\pagestyle{plain}

\def\e{{\sffamily e}}
\def\cs{\frac{1}{(2\pi\alpha')^2}}
\def\CV{{\cal{V}}}
\def\haf{{\frac{1}{2}}}
\def\tr{{\sffamily Tr}}
\def\"{\prime\prime}
\def\p{\partial}
\def\tphi{\tilde{\phi}}
\def\ttheta{\tilde{\theta}}
\def\a{\alpha}
\def\b{\beta}
\def\la{\lambda}
\def\barla{\bar{\lambda}}
\def\ep{\epsilon}
\def\hj{\hat j}
\def\hn{\hat n}
\def\bz{\bar{z}}
\def\zk{{\bf{Z}}_k}
\def\h1{\hspace{1cm}}
\def\dd{\Delta_{[N+2k] \times [2k]}}
\def\ddbar{\bar{\Delta}_{[2k] \times [N+2k]}}
\def\u{U_{[N+2k] \times [N]}}
\def\ubar{\bar{U}_{[N] \times [N+2k]}}
\def\goes{\rightarrow}
\def\goal{\alpha'\rightarrow 0}
\def\ads2{AdS_2 \times S^2}
\def\ola{\overline {\lambda}}
\def\oep{\overline {\epsilon}}
\def\SP{symplectic invariance\hspace{2mm}}
\def\SPT{symplectic invariant\hspace{2mm}}
\def\kt{K_3\times T^2}
\def\ff{{\cal F}}
\def\tff{\widetilde{{\cal F}}}


\vspace{3cm}

\title{\bf Minimal redefinition of the OSV ensemble }
\author{Shahrokh Parvizi\thanks{Email: parvizi@theory.ipm.ac.ir }, \hspace{3mm} Alireza
 Tavanfar\thanks{Email: art@ipm.ir }
\\
{\small {\em Institute for Studies in Theoretical Physics and
Mathematics (IPM),}} \\
{\small {\em P.O.Box 19395-5531, Tehran, Iran}}
\\
}
\maketitle

\begin{abstract}
\noindent In the interesting conjecture,
$\rm{Z_{BH}=|Z_{top}|^2}$,
 proposed by Ooguri, Strominger and
 Vafa (OSV),  the black hole ensemble is a \emph{mixed} ensemble and the resulting degeneracy
 of states, as obtained from the ensemble
  inverse-Laplace integration, suffers from prefactors which do not respect the
  \emph{electric-magnetic duality}. One idea to overcome this deficiency, as
  claimed recently,
  is imposing nontrivial measures for the ensemble sum.\\
  We address this problem and \emph{upon a redefinition of the OSV ensemble} whose variables
  are as numerous as the electric potentials,
  show that for restoring the symmetry \emph{no non-Euclidean measure is needful}. In detail,
  we rewrite the OSV free
  energy as a function of new variables which are combinations of the
  electric-potentials and the black hole charges.
  Subsequently the Legendre transformation which bridges between the entropy and the black
  hole free energy in terms of these variables, points to a generalized
  ensemble. In this context, we will consider \emph{all} the cases of relevance: \emph{small}
  and \emph{large} black holes, with
or without $D_6$-brane charge. For the case of vanishing
$D_6$-brane charge, the new ensemble is \emph{pure canonical} and
 the electric-magnetic duality is restored \emph{exactly}, leading to
proper results for the black hole degeneracy of states. For more
general cases, the construction still works well \emph{as far as}
the violation of the duality by the corresponding OSV result is
restricted to a \emph{prefactor}. In a concrete example we shall
show that for black holes with non-vanishing $D_6$-brane charge,
there are cases where the duality violation \emph{goes beyond this
restriction}, thus imposing non-trivial measures is
\emph{incapable} of restoring the duality. This observation
signals for a \emph{deeper modification} in the OSV proposal.

\end{abstract}

\vspace{1cm}

IPM/P-2005/059
\newpage
\Section{Introduction} The theory of \emph{topological strings} is
a beautiful, smart and powerful mathematics to probe various
aspects of the superstring theories. From the time of its
invention
\cite{witten:1988invention}\cite{Witten:1989ig}\cite{Witten:1991zz},
it has been an active field of new discoveries. The reader may
refer to \cite{Vafa:book} as an excellent textbook on the
topological strings, or consult
\cite{neit:2004rev}\cite{Vonk:2005yv}\cite{Marino:2004eq} as quick
introductions to the subject and \cite{neit:2004rev} for a nice
view to the recent developments.\\ On the other hand, one of the
most striking outcomes of the string theory has been the
microscopic derivation of the black hole entropy
\cite{vafa:1996be}. Recently Ooguri, Strominger and Vafa (OSV) in
\cite{Ooguri:2004zv} have proposed a deep and promising connection
between the topological free energy and the microstate ensemble of
the $4$-dimensional BPS Black Holes in the $CY_3$
compactifications of type-II strings. This conjecture roots from
earlier works on the black hole entropy
\cite{wald:1993for}\cite{moha:1999cor}\cite{LopesCardoso:1999cv}
which made a consistent generalization of the Bekenstein-Hawking
formula, incorporating the F-terms corrections to the supergravity
action as encoded in the prepotential. The evidence for the fact
that a generalization to the standard area-law formula for the
black hole entropy is needful in the context of string-M theory
had been provided in \cite{Maldacena:1997one}. A detailed review
on the results about the black hole entropy from the string theory
is \cite{mohaupt:2000}.\\ In \cite{Ooguri:2004zv}, OSV observe
that the Bekenstein-Hawking-Wald entropy of the black holes, as
derived in \cite{moha:1999cor}\cite{LopesCardoso:1999cv}, can be
\emph{re-expressed} as the \emph{Legendre} transformation of a
real function $\ff$, called the OSV free energy, with respect to
the half of its variables, \bea\label{bhw}\rm{
S(p^{\Lambda},q_{\Lambda})\; = \; \ff(p^{\Lambda},\phi^{\Lambda})+
q_{\Lambda}\phi^{\Lambda}\;\;\;;\;\;\;\ff(p^{\Lambda},\phi^{\Lambda})\equiv
- \pi \; \Im F(CX^\Lambda,C^2W^2=2^8)}\eea with \bea\label{at1}
\rm {CX^\Lambda \;=\;p^\Lambda\; + i \; \phi^\Lambda}\;\;\; \eea
as obtained from the magnetic part of the attractor equations, and
$\rm{F(X^\Lambda)=F(X^0,X^A)}$ being the full \emph{holomorphic}
prepotential. Independently, a very close relation is established
between the asymptotic expansion of the prepotential and the free
energy of topological strings
\cite{Bershadsky:1993cx}\cite{Antoniadis:1993ze}. In all the OSV
conjecture states that, given a BPS black hole of
electric-magnetic charge-multiplet $(p^\Lambda,q_\Lambda)$,
arising in $\rm{\mathcal{N}}$$\rm{=2}$ compactification of
$IIA(B)$ string theory on $CY_3=\cal{M}$, the topological
$A-\bar{A}$$(B-\bar{B})$ models on $\cal{M}$ define a mixed
ensemble as, \bea \label{OSV} \rm{Z_{mix} \equiv e^
{\ff(p^\Lambda,\phi^\Lambda)} = \sum_{q_\Lambda} \;\;
\Omega(p^\Lambda,q_\Lambda)\; e^{-\pi\; q_\Lambda\phi^\Lambda}}
\eea where the \emph{black hole free energy} is given by
\bea\label{FE} \rm{F_{BH}=\ff(p^\Lambda,\phi^\Lambda) = F_{top} +
\bar{F}_{top}} \eea with $\rm{\Omega(p^\Lambda,q_\Lambda)}$
proposed as (index) degeneracy of the black hole states and
$\rm{\phi^\Lambda}$ being the electric potential corresponding to
$\rm{q_\Lambda}$.\footnote{For a non-perturbative completion of
the original conjecture see \cite{Dijkgraaf:2005bp}, also there
are suggestions concerning the holomorphic anomaly
\cite{mina:2004ym}\cite{Verlinde:2004ck}\cite{Shih:2005he}.} This
proposal has been successfully tested in \cite{mina:2004ym} and
\cite{Dabholkar:2004yr} from different points of view and further
developed in \cite{Dijkgraaf:2005bp}. Remarkably
 using the OSV conjecture \cite{Dabholkar:2004yr} was able for the first time to derive
 from
 the macroscopic side, the \emph{exact} coefficient of the leading term in the asymptotic
 expansion of the degeneracy of states for the small black hole of type-IIA on $K_3\times T^2$,
 dual to Heterotic on $T^6$, where both the prepotential and the microscopic-degeneracy counting are
  known exactly. For a review lecture on \cite{Ooguri:2004zv} \cite{mina:2004ym} and
  \cite{Dijkgraaf:2005bp}, see \cite{Ooguri:2005lec}.\\
There are however examples in
\cite{Dabholkar:2005by}\cite{Dabholkar:2005rev} where a
\emph{naive} application of this conjecture sounds problematic.
For example \cite{Dabholkar:2005by} mainly focuses on
$\mathcal{N}$ $=4$ models where in some cases
 the string coupling is strong. Before concentrating on the concern of this paper,
 it is worth to point out that before any try to apply or to refine the proposal
 one may note that the OSV conjecture in form of \cite{Ooguri:2004zv}\cite{Dijkgraaf:2005bp}
  is a \emph{statement} about the \emph{full free energy} of the topological string theory,
  including all the perturbative quantum corrections as well as the non-perturbative
   contributions. Therefore regarding the fact that the non-perturbative completion of the
    topological free energy is not known so far, what the OSV conjecture provides is so a
     non-perturbative definition of the topological strings, parallel to the program of
     topological M-theory as proposed in \cite{Dijkgraaf:2004te} and \cite{Pestun:2005rp}.\\
A distinguished ambiguity in (\ref{OSV}) originates from the
electric-magnetic duality considerations, that is \emph{the
requirement of the invariance of $\Omega(p^I,q_I)$ under the
(relevant subgroup) of the symplectic transformations}. Actually
as concrete examples show
\cite{Dabholkar:2005by}\cite{Pioline:2005vi}\cite{Dabholkar:2005rev},
given the \emph{relevant} terms of the topological free energy,
the inverse Laplace transformation corresponding to
(\ref{OSV})\footnote{The sign $\doteq$ means `` equality " up to
an arbitrary constant which is independent of the black hole
charges.}, \bea\label{inversel}\rm{
\Omega(p^\Lambda,q^\Lambda)\;\;\doteq\;\;\int [d\phi^\Lambda] \;\;
e^{\ff\;+\;\pi\;q_\Lambda\phi^\Lambda}\;\equiv\;\int
[d\phi^\Lambda] \;\; e^{G}} \eea
 \emph{does} lead to the
results which are duality violating due to some unwanted
prefactors, for a highlighted example of which see the next
section . As we will argue later, this property is not restricted
to a specific subset of prepotentials but it is a \emph{general}
fact about (\ref{inversel}). Modifying this deficiency has been a
matter of some investigations. For example, as a try within an
independent and, by construction, duality invariant formulation
see \cite{Dewit:2005lec}, where the proposed modification works
for the dyonic black holes relatively well but it is not
applicable to the case of $\frac{1}{2}$BPS states for which the
proposed measure vanishes. Alternatively
\cite{Sen:2004dp}\cite{Sen:2005pu}\cite{Sen:2005ch} propose a
black hole ensemble which sums over a \emph{single}
duality-invariant charge, so by construction respects the expected
symmetries. The application of this approach, where the
\emph{non-holomorphic} corrections to the free energy should
necessarily be considered, is so far limited to $\frac{1}{2}$BPS
black holes. So there is not yet any overlap between these two
approaches. Another idea to improve the duality violations, being
closer to the OSV proposal, is assuming that the OSV free energy
\emph{intrinsically} lives in a \emph{curved} space, accordingly
(\ref{inversel}) is redefined via a \emph{non-Euclidean} measure
\cite{Dabholkar:2005rev}\cite{Shih:2005he},
\bea\label{inversegl}\rm{
\bar{\Omega}(p^\Lambda,q_\Lambda)\;\;\doteq\;\;\int
[d\phi^\Lambda]\;\sqrt{|g(p,\phi)|} \;\;
e^{\ff\;+\;\pi\;q_\Lambda\phi^\Lambda}}\;\;\;.\eea This measure
which differs from case to case is responsible
 for removing the mentioned prefactors. Along this path, \cite{Dabholkar:2005rev}
 calls for a
 ``deeper understanding of the integration measure implicit in (\ref{inversel})".\\
In this paper we focus on this problem, that is the duality
aspects of the OSV conjecture. We shall prove, at least
asymptotically, the converse of the statement which leads to
(\ref{inversegl}). That is, the \emph{curvature} of the space
where the OSV free energy is defined, asymptotically or exactly,
is \emph{zero}. To show this we shall follow a direct approach:
explicit construction of a proper $flat$ ensemble, based on the
OSV free energy, which is canonic in \emph{as many variables as}
the original OSV ensemble. In different examples, also within
general arguments, we observe that the ensemble leads to the
proper results for the black hole degeneracy of states. In fact
the construction works properly for \emph{all} types of the known
BPS black holes in $CY_3$ compactifications of type-IIA(B)
theories and to \emph{all} orders in the saddle point asymptotic
expansion of the inverse Laplace integral, as far as the duality
violation is restricted to a prefactor. This is done via a
\emph{simple} change of variables which being linear in the
electric-potentials, preserves the Legendre transformation from
the free energy to the entropy. The new variables have the
advantage that the corresponding ensemble is \emph{potentially}
protected against the duality violating prefactors. For the cases
with vanishing $D_6$-brane charge, where the prepotential takes a
simplified form and the construction is readable from the form of
the prepotential, the new ensemble is \emph{pure canonical} and
restores the electric-magnetic \emph{exactly}. The idea still
works quite well for the most general case, of course up to
\emph{nonholomorphic} corrections which are missed also in the
original OSV proposal. For the most general case, considering the
leading term of the saddle-point approximation, we obtain a
general constraint on the Jacobian from the new variables to the
old ones. This
 constraint-equation admits \emph{infinite} number of solutions, in principle all the solutions
 are equally well, but practically we can pick up the simplest choices.
  In fact the exact choice of these variables is not needed.
   The job they do is just removing the duality-violating prefactor of the
    corresponding OSV result, which could be done \emph{by hand} from the first.
    In this sense it is an \emph{existence} problem. \\ It is
    worth to ask about higher orders of the saddle-point
    approximation for the most general case. The question is whether the duality violation
    appears just as a prefactor in higher orders. Indeed, for the black holes with
    vanishing $\rm{D_6}$-brane charge the answer is positive to all orders of the asymptotic
    expansion, however we will
    show that in case of non-vanishing $\rm{D_6}$-brane charge, the violation can become
    more complicated than a simple prefactor. This implies that, in
    higher orders, a simple measure factor is not sufficient to
    restore the duality generally and deeper modifications/investigations
    are necessary. \\ Finally as an \emph{alternative},
    motivated by \emph{manifest} symplectic-invariant construction, we try a different
    ensemble which from the \emph{beginning} treats
    both of the magnetic and electric charges at the same footing.
    This \emph{enlarged} ensemble is related to the ensemble of section $4$ via an
    \emph{effective} integration over the \emph{magnetic potentials} and asymptotically
     reproduces
    the same results for the black hole degeneracy of states.\\
    The outline of the paper is as follows. In the next section, for the clearness
    of the main problem within a concrete example, we \emph{review} one result from
    \cite{Dabholkar:2005by}\cite{Dabholkar:2005rev}. In section
    3, we redefine the OSV ensemble for the
    large and small black
    holes in the absence of $D_6$-brane, respectively, and show how it
    leads to proper results for the degeneracy of states.
    In section 4 we lift the idea for the most general case of
    BPS black holes in the context of $CY_3$ compactifications of
    type-IIA(B) string theories. In section 5, we check out the procedure for higher
    orders in the saddle-point approximation and show that in some special cases
    the duality violation can
    not be fixed by a single prefactor.
    In section 6 we introduce an
    effective which asymptotically reproduces the same results of section 4.
    We end the paper with a conclusion.
\Section{Electric-magnetic duality in the OSV conjecture} Let us
start by addressing the duality invariance of the proposal of
\cite{Ooguri:2004zv}.
 To do that, we repeat swiftly a result from \cite{Dabholkar:2004yr}, \cite{Dabholkar:2005by}
and \cite{Dabholkar:2005rev}\footnote{This example is in the
context of $\mathcal{N}$$=4$, but for $\mathcal{N}$$=2$, where the
OSV conjecture is originally formulated, this is the case as well.
See for example the next section.}. The prepotential of type $II$
string theory on a $CY_3$ is given by the free energy of the
corresponding topological string theory as follows : \bea\rm{
F_{{\sffamily II}} = -\frac{i \pi}{2} F_{{\sffamily top}}
\;\;\;\;\;\;\;\;\; F_{{\sffamily top}}=F_{{\sffamily
top}}^{{\sffamily pert.}} +F_{{\sffamily top}}^{{\sffamily GW}}+
F_{{\sffamily top}}^{{\sffamily non-pert. \; completion}}} \eea
where \bea\rm{ F^{pert.}=\sum_h F_h W^{2h}\;}, \eea with $F_h$
being the genus-$h$ amplitude of the topological theory and $W^2$
includes the graviphoton field strength. For the case of IIA on
$\kt$, $F_{h>1}$ vanishes and the prepotential is given by,
\bea\rm{ F(X^I,W^2)= -\frac{1}{2} C_{ab}\frac{X^a X^b X^1}{X^0}
-\frac{W^2}{128\pi i}\log\Delta\left(e^{2\pi i
\frac{X^1}{X^0}}\right)}, \eea where $\rm{C_{ab}\equiv C_{1ab}}$
with $\rm{a,b= 2, \dots ,b_2}$, $\rm{b_2=23}$, and $\rm{C_{ABC}}$
being the intersection numbers of the $CY_3$. $\Delta$ stands for
the Dedekind function. Taking $\rm{p^0=0}$, the OSV free energy is
given by,
 \bea\label{free1}\rm{
\ff(p^\Lambda,\phi^\Lambda)=
\frac{\pi}{2\phi^0}C_{ab}(p^1\phi^a\phi^b + p^a\phi^b\phi^1 +
p^b\phi^a\phi^1 - p^a
 p^b p^1)-\log\left|\Delta\left(\e^{2\pi  \frac{p^1}{\phi^0}} \e^{2\pi i \frac{\phi^1}
 {\phi^0}}\right)\right|^2 }\;\;\;.
\eea For the small black hole we set $p^a=0$,
\bea\label{free3}\rm{ \ff(p^I,\phi^I) = \frac{1}{2}
C_{ab}\frac{\phi^a \phi^b p^1}{\phi^0} - \log(\;|\Delta
\;(e^{2\pi^2 \frac{p^1}{\phi^0}} e^{2\pi i
\frac{\phi^1}{\phi^0}}\;)\;|^2)\;\;\;} . \eea Accordingly the
inverse Laplace transformation, with the Euclidean measure,
calculates the degeneracy of states as, \bea
\rm{\Omega(p^1,\vec{q})\;\doteq\; \int \;[d\phi^0 d\phi^1 d\phi^a
\;] \e^{\ff+q_a\phi^a+q_0\phi^0+q_1\phi^1}} \eea \bea\rm{\doteq\;
\int dx d\theta \;(\frac{-x^{12}(p^1)^2} {\sqrt{\det C}} \;)
\frac{\e^{(\haf C^{ab}q_a q_b - p^1 q_0)x -p^1q_1 x\theta}}
\;|\Delta\;(\e^{-\frac{2\pi^2}{x}} \e^{2\pi i \theta}\;)\;|}\eea
where in the second line, we have taken the integration over
$\phi^a$'s and used the change of variables $x=-\phi^0/p^1$ and
$\theta=\phi^1/\phi^0$ with $d\phi^0d\phi^1=(p^1)^2 x dx d\theta$.
For the two-charge case, we set $q_A=0$ for $A \neq 0$. Defining
the T-duality invariant charge $N \equiv -p^1q_0$, the integrating
over $\phi^{a}$ gives, \bea \rm{ \Omega(p^1,q_0) \; \doteq \;
\frac{-(p^1)^2}{\sqrt{\det C}} \int dx x^{12}\e^{Nx} \int d\theta
\frac{1}{\left|\Delta\left(\e^{-\frac{2\pi^2}{x}} \e^{2\pi i
\theta}\right)\right|}} \eea
 In this form, beside the prefactor
$(p^1)^2$, everything is in a duality invariant form. The presence
of this $p$-dependent factor violates the duality invariance of
the degeneracy of states.

 In the large $N$ limit,
$|\Delta(q)|\approx |q|$, so that $\Omega$ approaches, \bea
\label{omega} \rm {\Omega(p^1,q_0)\doteq
\frac{-(p^1)^2}{\sqrt{\det\; C}}\;\hat{I}_{13}(4\pi \sqrt{N})}
\eea where \cite{Dabholkar:2005by}, \bea \rm{\hat{I}_{\nu}(Q)
\equiv \frac{(2\pi)^\nu}{i} \int d t \;t^{-\nu-1}
\e^{t+\frac{Q^2}{4 t}}}. \eea Note that at the above limit, the
genus 0 and 1 terms of prepotential read as, \bea \label{tree}
\rm{F_0(X^\Lambda)\;=\; -\frac{1}{2} C_{ab}\frac{X^1 X^a
X^b}{X^0}} \eea \bea \rm{F_1(X^\Lambda)\;=\;
-\frac{1}{64}\frac{X^1}{X^0}} \eea with the OSV free energy, being
\bea\label{free2}\rm{ \ff^{pert}_{IIA/\kt}\;=\;\frac{C_{ab}}{2}
\phi^a\phi^b \frac{p^1}{\phi^0}\;-\;4\;\pi^2 \frac{p^1}{\phi^0}}
\eea We mention that any attempt to remove the prefactor of
(\ref{omega}), and similar prefactors, should be done in a way to
keep the correct results for the entropy and degeneracy of states.
In the following sections we redefine the black hole ensemble to
do this job.
\Section{Redefinition of the OSV ensemble for a class of large and
small black holes} In this section we consider black holes with
vanishing $D_6$-brane charge, where $\rm{p^0=0}$ and the
prepotential takes a simplified form. We start with the case of
\emph{large} black holes . The perturbative OSV free energy for a
general $CY_3$ compactification in the large volume is given by,
\bea\label{large}\rm{
\ff=-\frac{\pi}{6}\frac{\hat{C}(p)}{\phi^{0}}+\frac{\pi}{2}\frac{C_{AB}
\phi^{A}\phi^{B}}{\phi^{0}}} \eea where \bea\rm{
C_{AB}(p)=C_{ABC}p^C\;\;\;,\;\;\;
C(p)=C_{ABC}p^Ap^Bp^C\;\;\;;\;\;\;\hat{C}(p)=C(p)+C_{2A}p^A} \eea
and the indices $A,B,...$ take the integers $1,...,h$. \\ Let us
change the variables of the free energy as follows,
\bea\label{nvl}\rm{
 \psi^{0} \equiv \frac{6}{\hat{C}(p)}\phi^0\;\;\;;\;\;\;\frac{C_{AB}(p)\phi^A\phi^B}
 {\phi^0} \equiv \frac{\psi^A \psi^A}{\psi^0}}
\eea where$\;\;\rm{\psi^A = M^{A}_{\;\;B}(p)
\phi^B}$.\footnote{$\rm{M^{A}_{\;\;B}(p)}$ is a transformation
matrix which first diagonalizes $\rm{C_{AB}(p)}$, then rescales
the
variables appropriately.}\\
The free energy is now redefined as
 a new function,
\bea\label{preinnew}\rm{
\ff\;=\;\grave{\ff}\;(\psi^0,\psi^A)\;=\;-\pi\frac{1}{\psi^0}+\;\frac{\pi}{2}
\frac{\psi^A \psi^B} {\psi^0}\;\;}. \eea Noting that \bea\rm{
 -\phi^0 \frac{\partial
\ff}{\partial \phi^0} = -\psi^0 \frac{\partial
\grave{\ff}}{\partial \psi^0}\;\;\;\;\;\;;\;\;\;\;-\phi^A
\frac{\partial \ff}{\partial \phi^A} = -\psi^A \frac{\partial
\grave{\ff}}{\partial \psi^A}} \eea and defining \bea\rm{ N \equiv
\frac{1}{6} \hat{C}(p)q_0 \;\;\;;\;\;\; N_A \equiv q_B
(M^{-1})^{B}_{\;\;A}} \eea the Legendre transformation reads as,
\bea\rm{ S(N,N_A)=\grave{\ff}- \pi \psi^0 \frac{\partial
\grave{\ff}}{\partial \psi^0}- \pi \psi^A \frac{\partial
\grave{\ff}}{\partial \psi^A}} \eea \bea \rm{N=-\frac{\partial
\grave{\ff}}{\partial \psi^0}\;\;\;;\;\;\;N_A=-\frac{\partial
\grave{\ff}}{\partial \psi^A}}\;\;. \eea Motivated by the above
observation we introduce the \emph{correct black hole ensemble to
use} as \bea\label{new} \rm{Z \equiv
e^{\grave{\ff}(\psi^0,\psi^A)} = \sum_{N} \sum_{N_A}
\tilde{\Omega}(N,N_A)\;e^{-\pi\psi^0 N-\pi\psi^A N_A}}\;\;. \eea
Note that the ensemble defined above, unlike the OSV ensemble, is
\emph{pure canonical}. Now the black hole degeneracy of states is
proposed to be \bea \rm{d(p,q)\;\;\; \dot{=}\;\;\;
\tilde{\Omega}(N,N_A)} \;\;\;.\eea
 Let us now consider (\ref{large}) and calculate the corresponding degeneracy of states
in the light of the definition (\ref{new}), \bea\rm{
\tilde{\Omega}(N,N_A) \; \dot{=} \; \int d\psi^0\; d\psi^A\;
e^{\grave{\ff}(\psi^0,\psi^A) +  \pi N \psi^0 + \pi N_A \psi^A }}
\eea The integral over the variables $\psi^A$'s is Gaussian and
yields \bea\rm{ \tilde{\Omega}(N,N_A) \; \dot{=} \;
\hat{N}^{-\frac{h}{2}-1} \int d\mu \; \mu^{\frac{h}{2}} e^{\mu -
\pi^2 \frac{\hat{N}}{\mu}}} \eea where \bea\rm{ \hat{N} \equiv N -
\frac{1}{2} N_A N_A} \eea A simple algebra shows that \bea
\rm{\hat{N} = \frac{1}{6} \hat{C}(p) \hat{q}_0 \;\;\;;\;\;\;
\hat{q}_0 \equiv q_0 - \frac{1}{2} C^{AB} q_A q_B} \eea where
$\rm{C^{AB}}$ denotes the inverse of $\rm{C_{AB}}$. Thus the
degeneracy of the large black hole states as calculated via the
ensemble (\ref{new}) is given by, \bea\label{pra} \rm{d(p,q)\;
\dot{=}\; \hat{I}_{\frac{h}{2}+1}(S\;)\;\;\;;\;\;\;S=2\pi\sqrt
{\frac{1}{6}{\hat{C}(p)\hat{q}_0}}}\;\;\;\eea This is indeed the
correct result. In contrast, the result by the standard OSV
ensemble receives a duality non-invariant prefactor of the form,
$|\det
C_{AB}(p)|^{-\frac{1}{2}}(\frac{\hat{C}(p)}{6})^{\frac{h}{2}+1}$
\cite{Dabholkar:2005by} \cite{Dabholkar:2005rev}.

Now we come to the case of \emph{small} black holes following the
same idea: rewriting the OSV free energy and the macroscopic
entropy in terms of new variables and new charges respectively
leads to a new ensemble for the black holes which by construction
is
 \emph{pure canonical}. For the interesting case of $K3$-fibrations with Heterotic duals,
of which the most well-known example being $\frac{IIA}{\kt} \equiv
\frac{Het.}{T^6}$,
 the OSV free energy is (\ref{free3}).
Defining \bea\label{nvs}\rm { \alpha \; \equiv \; -
\frac{\phi^0}{p^1}\;\;\;;\;\;\; \beta \; \equiv
\frac{\phi^1}{p^1}} \eea and \bea \rm{N \; \equiv \;
-p^1q_0\;\;\;;\;\;\; N^{\prime} \; \equiv \; -p^1q_1\;\;\;. }\eea
We obtain \bea\rm{
\grave{\ff}(\alpha,\beta,\phi^a)\;=\;-\frac{C_{ab}\phi^a\phi^b}{2\alpha}
-2 \log | \Delta(e^{\frac{2\pi^2}{\alpha}}e^{\frac{-2\pi i
\beta}{\alpha}})|} \eea where \bea\rm{
S(N,N^{\prime},q^a)\;=\;\grave{\ff}- \pi \alpha \frac{\partial
\grave{\ff}}{\partial \alpha}
 - \pi \beta \frac{\partial \grave{\ff}}{\partial \beta} - \pi \phi^a
  \frac{\partial \grave{\ff}}{\partial \phi^a}}
\eea and \bea \rm{N = -\frac{\partial \grave{\ff}}{\partial
\alpha}\;\;\;;\;\;\; N^{\prime} = -\frac{\partial
\grave{\ff}}{\partial \beta}\;\;\;;\;\;\;q_a= -\frac{\partial
\grave{\ff}}{\partial \phi^a}} \eea Subsequently we define
\bea\label{nvse}\rm{ \grave{Z} \equiv e^{\grave{\ff}(\alpha,
\beta, \phi^a)} = \sum_{N} \sum_{N^{\prime}}
 \sum_{q_a} \grave{\Omega}(N,N^{\prime},q_a)\;
 e^{-\pi\alpha N-\pi\beta N^{\prime}-\pi\phi^a q_a}}
\eea
as the correct black hole ensemble.\\
The corresponding degeneracy of states is so given by, \bea\rm{
\grave{\Omega}(N,N^{\prime},q_a) \; \dot{=} \; \int d\alpha\;
d\beta\;
 d\phi^a \;\;\; \frac{e^{-\frac{C_{ab}\phi^a\phi^b}{2\alpha} +
  \pi N \alpha + \pi N^{\prime} \beta + \pi q_a \phi^a}}{|\Delta(e^{\frac{2\pi^2}{\alpha}}
  e^{\frac{-2\pi i \beta}{\alpha}})|^2}}
\eea Doing the Gaussian integration and changing the variables of
the remaining integrals from $(\alpha,\beta)$ to $(\alpha,\theta
\equiv - \frac{\beta}{\alpha})$, we obtain \bea\rm{
\grave{\Omega}(N,N^{\prime},q_a) \; \dot{=} \; \int d\alpha\;
\alpha^{12}\; e^{\pi \alpha \hat{N}} \;\int \; d\theta \frac{
e^{-\pi N^{\prime}\theta\alpha}}
{|\Delta(e^{\frac{2\pi^2}{\alpha}}e^{2\pi i \theta})|^2}} \eea
where \bea\label{inte}\rm{\hat{N} \; \equiv \; N + \frac{1}{2} q_a
C^{ab} q_b = \frac{1}{2} Q_{e}^{2}} \eea For the 2-charge case
where $\hat{N}=N$ and $N^{\prime}=0$, taking the large-volume
limit , for which $|\Delta(x)| \approx |x|$, simplifies
(\ref{inte}) as \bea\label{result1} \rm{\tilde{\Omega}(N) \;
\dot{=} \; \int d\alpha\; \alpha^{12}\; e^{\pi \alpha \hat{N}+
\frac{4\pi}{\alpha}}\;\;=\;\; \hat{I}_{13}(4\pi
\sqrt{\hat{N}}\;)}\eea which is the well known result for the
degeneracy of states, as known from its duality with Heterotic
string theory by direct counting of the DH fundamental stringy
states
\cite{Dabholkar:2004yr}.\\
\Section{Black holes with non-vanishing $D_6$-brane charge:
Saddle-point evaluation } In this section we concern the most
general case, $p^0 \neq 0$, and as the
 central idea follow what we did before. We introduce a generalized OSV ensemble as,
\bea\label{new0}\rm{ \grave{Z} \equiv e^
{\grave{\ff}(p^\Lambda,\psi^\Lambda)} = \sum_{\grave{q}_\Lambda}
 \;\; \grave{\Omega}(p^\Lambda,\grave{q}_\Lambda)\; e^{- \pi \grave{q}_\Lambda
 \psi^\Lambda}}
\eea \bea\label{new1}\rm{
\grave{\ff}(p^\Lambda,\psi^\Lambda)\;\;=\;\;
\ff(p^\Lambda,\phi^\Lambda)} \eea where the new variables
$\psi^\Lambda$ which preserve the entropy-defining Legendre
 transformation,
\bea\label{new3}\rm{ S(p^\Lambda,\grave{q}_\Lambda) = \grave{\ff}
+ \pi \grave{q}_\Lambda \psi^\lambda}\;\;\;. \eea are defined from
the electric potentials via the \emph{linear} relations
\footnote{The sign $*$ over or below a quantity stands for the
saddle-point value of that quantity.}, \bea\label{new2}\rm{
\psi^\Lambda = U^{\Lambda}_{\;\;\Gamma}(p^\Sigma,
\phi_{*}^{\Sigma})\; \phi^\Gamma =
 V^{\Lambda}_{\;\;\Gamma}(p^\Sigma, q_\Sigma) \; \phi^\Gamma}
\eea The fact that unlike the previous cases, here we let the
matrix $U$ to
 depend on $\phi_{*}^\Lambda$ as well as $p^{\Lambda}$, roots in the
  observation that for the cases with non-vanishing $D_6$-brane charge,
   the OSV prefactors are functions of both the magnetic and electric charges
   \cite{Pioline:2005vi}\cite{Dabholkar:2005rev}.\\
We now \emph{require} that the corresponding inverse Laplace
integration as defined via the Euclidean measure,
\bea\label{new4}\rm{
\grave{\Omega}(p^\Lambda,\grave{q}_\Lambda)\;\;\dot{=}\;\;\int
[d\psi^\Lambda]\;\; e^{\grave{\ff}+ \pi \grave{q}_\Lambda
\psi^\Lambda}} \eea be \emph{duality invariant}, exactly or at
least at the saddle-point approximation. Subsequently the black
hole degeneracy of states is recognized as, \bea\label{new5}
\rm{d(p^\Lambda,q_\Lambda)\;\; \dot{=}\;\;
\grave{\Omega}(p^\Lambda,\grave{q}_\Lambda)} \eea where as implied
by (\ref{new2}), \bea\label{new5}\rm{ \grave{q}_\Lambda = q_\Gamma
\;\;(V^{-1})^{\Gamma}_{\;\;\Lambda}\;\;\;}. \eea
\\From (\ref{new1}) and (\ref{new2}), the relation between the
ensemble (\ref{new0}) and (\ref{OSV}) reads from \footnote{Since
the matrices $U$ and $V$ differ from each other just by a change
of variable $q\; \leftrightarrow\; \phi_{*}\;$, hereafter both of
them are denoted as $V$.}, \bea\label{rela}\rm{ \grave{\Omega}\;
\doteq\; \int [d\phi^\Lambda]\; J(\psi\rightarrow\phi)\; e^{\ff +
\pi q_\Lambda \phi^\Lambda} \;= \; \det[V]\;\int [d\phi^\Lambda]\;
e^{\ff + \pi q_\Lambda \phi^\Lambda}\;=\; \det[V]\;\Omega\;\;}.
\eea
 Obviously
 a proper choice of the Jacobian matrix $V$ can remove any
unwanted prefactor as obtained by the evaluation of
(\ref{inversel}). We so consider the cases where the result by
(\ref{inversel}) is duality-violating \emph{only} by a (single)
prefactor besides the proper result\footnote{Whether this is a
general property of the OSV results, is a question which we
address in the next section.}. Then for the duality invariance in
the saddle-point asymptotic expansion of (\ref{rela}), a
$constraint$-equation on $U^{\Lambda}_{\;\;\Gamma}$ is formed as a
necessary and sufficient condition. The exact form of this
constraint may depends on the order of the saddle-point
approximation. Here we write it for the first order. The
saddle-point evaluation of (\ref{rela}) results at, \bea\rm{
\grave{\Omega}^{*}\;\;=\;\;\det(V) \;\Omega^{*} }\;\;\;. \eea Thus
at the first order, \bea \rm{\grave{\Omega}^{*} \;\;\dot{=}\;\;
\det[V]\; \det[H_{*}]^{-\frac{1}{2}}\; e^S} \eea \bea\rm{
H_{\Lambda\Gamma}\;\; \equiv\;\; \frac{\partial^2 \ff}{\partial
\phi^{\Lambda} \partial \phi^{\Gamma}}}. \eea Remembering that
what we want to do is just to get rid of the unwanted prefactor
appearing in $\Omega$ and to get a proper result, we define,
\bea\rm{ \det[H_{*}]\;\; \equiv \;\;S^m \det[H^{\prime}_{*}]} \eea
where $m$ denotes a convenient power of the entropy which is
factorized out from the Hessian, so that
$\rm{det[H^{\prime}_{*}]}$ is the pure duality violating
prefactor. To remove this prefactor by the matrix $\rm{V}$, we
arrive at the constraint, \bea\label{cons1}\rm{
\det[V^2]\;\doteq\;\det[H^{\prime}_{*}]}\;\;\;. \eea The equation
(\ref{cons1}) admits \emph{infinite} number of solutions, any
solution of which is, in principle, as good as another ones. One
choice is taking $V\;=\;v\;{\bf 1}$ where
,\bea\label{cons2}\rm{v\;\;\doteq\;\;
(\det[H^{\prime}_{*}])^{\;\frac{1}{2h+2}}}\;\;\;. \eea As a
different choice for $V$ we can take a diagonal matrix with
elements $v_\Lambda$ such that, \bea\label{cons3}\rm{
v_\Lambda\;\doteq\;\sqrt{h^{\prime}_{\Lambda}}} \eea where
\bea\rm{ \prod_{\Lambda}
h^{\prime}_{\Lambda}\;=\;\det[H^{\prime}_{*}]\;\;\;.}\eea
Equivalently, we can solve the the constraint (\ref{cons1}) by
defining the new ensemble to be \bea\label{pos}\rm{\hat{X}^0
\equiv \det[V]\;X^0\;\;\;;\;\;\; \hat{X}^A \equiv X^A\;\;\;.}\eea
This solution \emph{might be} of particular physical interest
since $\rm{X^0}$ controls the coupling of topological strings via
the relation \bea\label{topcpl}\rm{g\;=\;\frac{4 \pi
i}{X^0}\;\;\;.}\eea In the conclusion we shall briefly speculate
on a possible physical interpretation of the ensemble (\ref{pos}).\\
As a specific example in the sector $ p^0 \neq 0 \;$, let us
consider the case of a large black hole in type-IIA on a $\rm{K}3$
fibered $CY_3$, for which
$C_{abc}=C_{2A}=0\;\;;\;\;{a,b}\in{2,...,h}$, as discussed in
\cite{Dabholkar:2005rev}\cite{Pioline:2005vi}. The prepotential is
\bea\label{prea}\rm{
 \ff = \frac{1}{2}\;[\;\Im(t^1)\;C_{ab}\;(p^a p^b
- \phi^a \phi^b)\;+\;2\;\Re(t^1)\; p^a q_a\;]} \eea
 with
$t^A=\frac{X^{A}}{X^{0}}$ being flat coordinates for the
K$\ddot{a}$hler moduli. The Hessian-determinant of (\ref{prea})
at the attractor point is evaluated as, \bea\rm{
\det[H_{*}]\;\doteq\;B^2\;(C_{ab} p^a p^b - 2 p^0 q_1
)^{\frac{h}{2}-1}
\;S^{-\frac{h}{2}-2}\;\equiv\;S^{-\frac{h}{2}-2}\;
\det[H^{\prime}_{*}] }\eea
 where
 \bea \rm{B \;=\;
\sqrt{(C_{ab}p^ap^b-2 p^0q_1)[(p^1)^2C_{ab}p^a
p^b+(p^0)^2C^{ab}q_aq_b-2p^0p^1p^a q_a]}
\;;\;C^{an}C_{nb}\;=\;\delta^{a}_{b}}\;\;\;. \eea
 Regarding (\ref{cons1}), the new variables are defined via the constraint
\bea\rm{ \det[V]\;\doteq\;B\;(C_{ab} p^a p^b - 2 p^0 q_1
)^{\frac{h-2}{4}}} \eea
from which, in case of interest, one can exactly specify a new set of variables.\\
Finally we want to show how in a simple way the variables of
(\ref{preinnew}) can be deduced from (\ref{cons3}).  The attractor
point of the prepotential (\ref{large}) is \bea\rm{
(\phi_{*}^0)^2\;=\;-\frac{1}{6}\frac{\hat{C}(p)}{\hat{q}_{0}}
\;\;\;;\;\;\;\phi_{*}^A\;=\;-C^{AB}(p)\;q_{B}\;\phi_{*}^0} \eea so
that
 \bea\rm{
\det[H_{*}];\doteq\;\det[\;C_{AB}(p)\;]\;\frac{\hat{C}(p)}{\phi_{*}^{0}}\;\;\;
\Rightarrow\;\;\;\det[H^{\prime}_{*}]\;\doteq\;\det[\;\frac{C_{AB}(p)}{\hat{C}(p)}\;]\;
\hat{C}^{-2}(p)}
 \eea
 and then from (\ref{cons3}) we obtain
 \bea\rm{
v^0 \; \doteq \; \frac{1}{\hat{C}(p)} \;\;\; ; \;\;\; v^A \;
\doteq \;
[\;\frac{\omega_{A}(p)}{\hat{C}(p)}\;]^{\frac{1}{2}}}\;\;\;.
 \eea
with $\omega_A(p)$ being the eigenvalues of $C_{AB}(p)$. Thus for
the $\psi $ variables we obtain \bea\label{again}\rm{
 \psi^0 \; \doteq \; \frac{\phi^0}{\hat{C}(p)} \;\;\; ; \;\;\; \frac{C_{AB}(p)\phi^A\phi^B}
 {\phi^0} \;=\; \frac{\omega_{A}(p)\hat{\phi}^A\hat{\phi}^A}{\phi^0}\; \doteq \;
 \frac{\psi^A \psi^A}{\psi^0}}
\eea where $\hat{\phi}^A$ form a diagonal basis for $C_{AB}(p)$.
The result (\ref{again}) is in accordance with (\ref{nvl}).
\Section{Is a major modification needful?} The electric-magnetic
duality is restored by the ensemble (\ref{new0}), \emph{as long
as} the result by (\ref{inversel}) admits the general form
\bea\label{form}\rm{\Omega(p,q)\;\doteq\; N(p,q)\;
f(S)\;\;\;.}\eea We naturally ask if violation of the duality in
the results of (\ref{inversel}) is restricted \emph{only} to a
prefactor, $\rm{N(p,q)}$. Of course for vanishing
$\rm{D_{6}}$-brane charge, this is always true, at least for the
large $CY_3$-volume limit. In these cases the prepotential is
quadratic with respect to $\rm{\phi^A}$, so the exact evaluation
of (\ref{inversel}) results at the form (\ref{form}) with
$\rm{f(S)}$ being a modified Bessel function. Now we come to the
cases for which $\rm{p^0 \neq 0}$. The prepotential, at the genus
zero and one terms, is of the general form \bea\label{nice0}
\rm{\ff\;=\;\frac{p^0}{(\phi^0)^2 + (p^0)^2}\;[\;E(\phi)-3
E_{AB}(\phi) p^A p^B - E_{2A}
\phi^A\;]}\eea$$+\;\rm{\frac{\phi^0}{(\phi^0)^2 +
(p^0)^2}\;[\;E(p)-3 E_{AB}(p) \phi^A \phi^B + E_{2A} p^A\;]}$$
where
$$\rm{E(z)\equiv E_{ABC}\; z^A z^B
z^c\;\;\;;\;\;\;\rm{E_{AB}(z)\equiv E_{ABC}\; z^c}}$$ with
$$\rm{E_{ABC} \equiv -\frac{\pi}{6}\;C_{ABC}\;;\;\;E_{2A} \equiv -\frac{\pi}{6}\;C_{2A}\;.}$$
To address the mentioned question for the case of non-vanishing
$\rm{p^0}$, we evaluate the OSV result for the black hole
degeneracy of states in concrete examples, through incorporating
\emph{all} the terms which appear in the saddle-point asymptotic
expansion of the integral (\ref{inversel}),
\bea\label{nice1}\rm{G\;=\;S + \sum_{n = 2 }^{\infty}\;
\frac{1}{n!} \;
H^{*}_{\Lambda_{1}...\Lambda_{n}}\;\eta^{\Lambda_{1}}...\eta^{\Lambda_{n}}}\eea
$$\rm{H_{\Lambda_{1}...\Lambda_{n}} \equiv \frac{\p^n \ff}{\p \phi^{\Lambda_{1}}...\p \phi^{\Lambda_{n}}}}$$
$$\eta^\Lambda \equiv \phi^\Lambda - \phi_{*}^{\Lambda} $$ for the two
\emph{extreme} limits: $\rm{p^0 \rightarrow \infty}$ and $\rm{p^0
\rightarrow 0}$.
\subsection{\underline{An example in the Large-$\rm{p^0}$ limit}}
The prepotential (\ref{nice0}) is expanded over
$\rm{\frac{\phi^{0}}{p^0}}$ as
$$\rm{\ff\;=\;\frac{1}{p^0}\;\{\;[\;E(\phi)-3 E_{AB}(\phi) p^A p^B -
E_{2A} \phi^A\;]} $$ $$
\;\;\;\;\;\;\;\;\;\;\;\;\;\;\;\;\;\;\;\;\;\;\;
\;\;\;\;\;\;\;\;\;\;\;\;\;\;\;\;\;\;\;\;\;\;\;+\;\rm{\frac{\phi^0}{p^0}\;[\;E(p)-3
E_{AB}(p) \phi^A \phi^B + E_{2A}
p^A\;]}\;+\;O\rm{(\frac{\phi^{0}}{p^0})^2}\;\}\;\;\;.$$ We require
that the contribution to (\ref{inversel}) from the regions of
integration far away from the attractor point is not major, and so
ignoring terms of higher orders in the above expansion is
satisfactory \emph{if} $O\rm{(|\frac{\phi_*^{0}}{p^0}|)}\ll1$. As
we will see this would be the case for
$\rm{|\frac{E(p)}{p^0}|\gg1}$ .\\
For the limit under consideration, the prepotential is linear with
respect to $\rm{\phi^0}$, so (\ref{nice1}) is easily evaluated as
$$\rm{G(\eta)\;=\;S\;-\;\frac{6}{(p^0)^2}\;E_{AB}(p)p^B\eta^0\eta^A\;}$$$$\;\;\;\;\;\;\;\;
\;\;\;\;\;\;\;\;\;\;\;\;\;\;\;\;\;\;\;\; +\;\rm{\frac{3}{p^0}
\;[\;E_{AB}(\phi_*)-\;\frac{\phi_*^{0}}{p^0}\;E_{AB}(p)\;
-\;\frac{\eta^0}{p^0}\;E_{AB}(p)\;]\;\eta^A\eta^B\;+\;\frac{1}{p^0}\;
E_{ABC}\eta^A\eta^B\eta^C\;}\;$$$$\rm{\equiv\;S\;+\;g_A
\eta^A\;+\;\frac{1}{2}\;g_{AB}\eta^A\eta^B\;+\;g_{ABC}\eta^A\eta^B\eta^C
\;}$$ Accordingly (\ref{inversel}) leads to
$$\rm{\Omega\;\doteq\;e^S\; \int d\eta^0 \;
\int\; [d\eta^A]\; e^{g_A\eta^A+\frac{1}{2}\;g_{AB}\eta^A
\eta^B+g_{ABC}\eta A \eta^B \eta^C}}$$$$\rm{=\;e^S\; \int d\eta^0
\; \int\; [d\eta^A]\; e^{g_Ax^A+\frac{1}{2}\;g_{AB}\eta^A \eta^B}
[\;1+\;g_{ABC}\eta^A \eta^B
\eta^C\;+\;O(\frac{1}{(p^0)^2})\;]\;\;\;.}$$ Now as a specific
example of the case under consideration, we consider a charge
configuration for which
$$\rm{\phi_*^A\;=\;p^A\;\;\;;\;\;\;\phi_*^0\;=0}$$ which is in line with the attractor
equations for
$$\rm{q_A\;=\;\frac{1}{p^0}\;E_{2A}}$$
$$\rm{q_0\;=\;\frac{1}{(p^0)^2}\;[\;2\;E(p)\;-\;E_{2A}p^A\;]}\;\;\;$$
with the entropy obtained as
$$\rm{S\;=\;-2\;\frac{E(p)}{p^0}}\;\;\;.$$
Given this ansatz, (\ref{inversel}) is given by
$$\Omega\;\ddot{=}\;\rm{\;e^S\; \int d\eta^0\;[d\eta^A]\;
exp\{\;\frac{1}{2}\;[\;\frac{6}{p^0}\;(1-\frac{\eta^0}{p^0})\;E_{AB}(p)\;]\;\eta^A
\eta^B-[\frac{6}{p^0}\;\frac{\eta^0}{p^0}\;E_{AB}(p)p^B\;]\eta^A\;\}\;
[\;1+\frac{1}{p^0}\;E(\eta)\;]\;\;\;}$$$$\rm{\equiv\;\Omega_0\;+\;\Omega_1}$$where
$\rm{\ddot{=}}$ differs from $\rm{\dot{=}}$ by higher order
corrections. A simple algebra leads to the result
$$\rm{\Omega_1\;\doteq\;\frac{(p^0)^{\frac{h}{2}+1}}{\sqrt{\det[E_{AB}]}}\;e^{-2S}
\;I_{\frac{h}{2}-1}(3S)}$$
$$\rm{\Omega_2\;\doteq\;\frac{(p^0)^{\frac{h}{2}+1}}{\sqrt{\det[E_{AB}]}}\;e^{-2S}
\;\int\;dt\;
t^{-\frac{h}{2}}\;[\;(1+\frac{1}{t})^3\;\frac{S}{2}\;-\;\frac{h}{2}\;\frac{1-t}{t^2}\;]
\;e^{\frac{3}{2}(t+\frac{1}{t})S}\;\;}$$ so in this example
$\rm{\Omega}$ is of the form (\ref{form}).
\subsection{\underline{ The infinitesimal-$\rm{p^0}$ limit}}
Here considering black holes with infinitesimal $\rm{D_6}$-brane
charge, $\rm{|p^0|\ll 1}$, we evaluate the OSV degeneracy of
states up to $O\rm{(|p^0|^2)}$. To do that, we do not restrict
ourselves to any finite order of the saddle-point approximation
and sum over \emph{all} the terms of the series (\ref{nice1}).
However we still assume that the integral (\ref{inversel}) is
essentially localized around the saddle-point of the integrand, so
that in the power expansion of the prepotential (\ref{nice0}) over
$\rm{\frac{p^0}{\phi^0}}$, we can ignore the terms
$O\rm{(\frac{p^{0}}{\phi^{0}})^2}$ as far as $\rm{\kappa \equiv
\frac{p^0}{\phi_{*}^0}\ll 1}$. As the attractor-point equations
will show us below, this assumption is \emph{consistent} with the infinitesimal-$\rm{p^0}$ limit. \\
The prepotential (\ref{nice0}), up to
$O\rm{(\frac{p^{0}}{\phi^{0}})^2}$, reads as\bea\label{nice3}
\rm{\ff\;=\;\frac{1}{\phi^0}\;\{\;\frac{p^0}{\phi^0}\;[\;E(\phi)-3
E_{AB}(\phi) p^A p^B - E_{2A} \phi^A\;]\;}\eea
$$\rm{\;\;\;\;+\;[\;E(p)-3 E_{AB}(p) \phi^A \phi^B + E_{2A}
p^A\;]\;\}}$$$$\rm{\equiv\;\rm{p^0\frac{K}{(\phi^0)^2}+\frac{L}{\phi^0}}\;\;\;.}$$
Regarding that the prepotential is a polynomial of degree three
with respect to $\rm{\phi^A}$'s, we can explicitly evaluate the
sum (\ref{nice1}). Using
$$\rm{H^{*}_{(n)}\;=\;(-1)^n[\;(n+1)!\;p^0\frac{K^{*}}{(\phi_{*}^0)^{n+2}}+\;n!\;\frac{L^{*}}
{(\phi_{*}^0)^{n+1}}\;]}$$
$$\rm{H^{*}_{A(n)}\;=\;(-1)^n[\;(n+1)!\;p^0\frac{K^{*}_{A}}{(\phi_{*}^0)^{n+2}}+\;n!\;
\frac{L^{*}_{A}}{(\phi_{*}^0)^{n+1}}\;]}$$
$$\rm{H^{*}_{AB(n)}\;=\;(-1)^n[\;(n+1)!\;p^0\frac{K^{*}_{AB}}{(\phi_{*}^0)^{n+2}}+\;
n!\;\frac{L^{*}_{AB}}{(\phi_{*}^0)^{n+1}}\;]}$$
$$\rm{H^{*}_{ABC(n)}\;=\;(-1)^n[\;(n+1)!\;p^0\frac{K^{*}_{ABC}}{(\phi_{*}^0)
^{n+2}}+\;n!\;\frac{L^{*}_{ABC}}{(\phi_{*}^0)^{n+1}}\;]}$$ where
$$\{K,L\}_{\{A_1...A_m\}} \equiv \frac{\p \{K,L\}}{\p \{A_1...A_m\}}$$
and the index $\rm{(n)}$ denotes the number of derivatives with
respect to $\phi^0$. Accordingly $\rm{G(\eta)}$ is evaluated as
$$\rm{G(\eta)\;=\;S\;+\;(\eta^0)^2\;[\;\sum_{0}^{\infty}\frac{(\eta^0)^n}{(n+2)!}
H^{*}_{(n+2)}\;]\;+\;\eta^0\eta^A
\;[\;\sum_{0}^{\infty}\frac{(\eta^0)^n}{(n+1)!}
H^{*}_{A(n+1)}\;]}$$
$$\;\;\;\;\;\;\;\;\rm{+\frac{1}{2}\;\eta^A\eta^B
\;[\;\sum_{0}^{\infty}\frac{(\eta^0)^n}{n!}
H^{*}_{AB(n)}\;]\;+\frac{1}{6}\;\eta^A\eta^B\eta^C
\;[\;\sum_{0}^{\infty}\frac{(\eta^0)^n}{n!} H^{*}_{ABC(n)}\;]}$$
$$\rm{=\;S\;+\;\frac{(\eta^0)^2}{(\phi_{*}^{0})^2}\;[\;\frac{p^0K^*}{(\phi_{*}^0)^2}
\;R_{3}(\frac{\eta^0}{\phi_{*}^0})+\;\frac{L^*}{\phi_{*}^0}
\;R(\frac{\eta^0}{\phi_{*}^0})\;]-\;\;\frac{\eta^0\eta^A}{(\phi_{*}^{0})^2}\;
[\;\frac{p^0K_{A}^*}{\phi_{*}^0}
\;R_{2}(\frac{\eta^0}{\phi_{*}^0})+\;L_{A}^{*}\;R(\frac{\eta^0}{\phi_{*}^0})\;]}$$
$$+\rm{\;\;\frac{1}{2}\;\frac{\eta^A\eta^B}{(\phi_{*}^{0})^2}\;
[\;p^0K_{AB}^*\;R_{1}(\frac{\eta^0}{\phi_{*}^0})+\;\phi_{*}^0L_{AB}^{*}
\;R(\frac{\eta^0}{\phi_{*}^0})\;]\;+\;\frac{1}{6}\;\frac{\eta^A\eta^B\eta^C}{(\phi_{*}^{0})^3}
\;p^0\phi_{*}^0K_{ABC}^*\;R_{1}(\frac{\eta^0}{\phi_{*}^0})}$$
where $$\rm{R(z)\;\equiv\;\sum_{0}^{\infty}(-1)^n\;
z^n\;=\;\frac{1}{1+z}\;\;;\;\;R_{m}(z)\;\equiv\;\sum_{0}^{\infty}(-1)^n\;
(n+m)\;z^n\;=\;\frac{(m-1)\;z+m}{(1+z)^2}}\;\;.$$
Defining$$\rm{x\;\equiv\;\frac{\eta^0}{\phi_{*}^0}\;\;\;;\;\;\;x^A\;\equiv\;
\frac{\eta^A}{\phi_{*}^0}}$$together with
$$\rm{g_{ABC}\;\equiv\;\kappa\frac{(\phi_{*}^0)^2}{(1+x)^2}\;E_{ABC}}$$
$$\rm{g_{AB}\;\equiv\;\frac{6\phi_{*}^0}{(1+x)^2}\;[\;\kappa\;
E_{AB}(\phi_*)-\;(1+x)\;E_{AB}(p)\;]}$$
$$\rm{}$$$$\rm{g_{A}\;\equiv\;-\frac{3\;x}{(1+x)^2}\;[\;\kappa(x+2)\;\{\;E_{AB}(\phi_*)
\phi_{*}^B-E_{AB}(p)p^B\;\}-2\;(1+x)\;E_{AB}(p)\phi_{*}^B-\kappa\;\frac{x+2}{3}\;E_{2A}\;]}$$
$$\rm{g\;\equiv\;\frac{x^2}{(1+x)^2}\frac{1}{\phi_*^0}\;[\;\{\;E(p)
(1+x)+\kappa\;E(\phi_*)(3+2\;x)\;\}}$$$$\;\;\;\;\;\;\;\;\;\;\;
\;\;\;\;\;\;\;\;\;-\rm{3\;\{(1+x)\;E_{AB}(p)\phi_{*}^A\phi_{*}^B+\kappa\;
(3+2\;x)E_{AB}(\phi_*)p^Ap^B\;\}}$$+$$\;\;+\;\rm{\{\;(1+x)\;E_{2A}p^A-\kappa
(3+2\;x)\;E_{2A}\phi_{*}^A\;\}\;]}$$ we obtain
$$\rm{G(\eta)\;=\;S + g + g_A\;x^A + \frac{1}{2}\;g_{AB}\;x^A x^B + g_{ABC}\;x^A x^B
x^C}\;\;\;.$$ Subsequently, (\ref{inversel}) is given by
$$\rm{\Omega\;\doteq\;e^S\; (\phi_{*}^0)^{h+1}\;\int dx \; e^g \int\;
[dx^A]\; e^{g_Ax^A+\frac{1}{2}\;g_{AB}x^A x^B+g_{ABC}x^A x^B
x^C}}$$$$\rm{=\;e^S\; (\phi_{*}^0)^{h+1}\;\int dx \; e^g \int\;
[dx^A]\; e^{g_Ax^A+\frac{1}{2}\;g_{AB}x^A x^B} [\;1+\;g_{ABC}x^A
x^Bx^C\;+\;O(\kappa^2)\;]}$$$$\rm{\ddot{=}\;e^S\;
(\phi_{*}^0)^{h+1}\;\int dx \; \frac{1}{\sqrt{\det[g_{AB}]}}
\;e^g\; (1+g_{ABC}\frac{\p^3}{\p g_C \p g_B \p
g_A})\;e^{-\frac{1}{2}\;g_{A}g^{AB}g_B}}$$
$$\rm{\ddot{=}\;e^S\;
(\phi_{*}^0)^{h+1}\;\int dx \; \frac{1}{\sqrt{\det[g_{AB}]}}\;
e^{T_{1}} \;(1+\;T_{2}\;+\;T_{3})}$$ where
$$\rm{T_{1}\;=\;g - \frac{1}{2}\;g_{A}\;g^{AB}\;g_{B}}$$
$$\rm{T_{2}\;=\;3 \;g_{ABC}\;g^{AB}\;g^{CV}\;g_{V}}$$
$$\rm{T_{3}\;=-\;g_{ABC}\;g^{AM}\;g^{BN}\;g^{CV}\;g_{M}\;g_{N}\;g_{V}}$$
and the sign `$\rm{\ddot{=}}$' meaning `$\rm{\dot{=}}$' up to
$O(\rm{\kappa^2})$. Now to proceed further and check whether the
above result is of the form (\ref{form}),
 we need to know some concrete information about how $\rm{g^{AB}}$ and
 $\rm{S}$ depend on $\rm{E_{ABC}}$, $\rm{E_{2A}}$ and the black hole charges.
In that respect, a helpful choice for us is to set
$$\rm{\phi_{*}^A\;=\;p^A\;\;\;\;}$$
according to which, the attractor equations is equivalent with
$$\rm{q_A\;=\;\frac{6}{\phi_{*}^0}\;p_A+\;\frac{p^0}{(\phi_{*}^0)^2}\;E_{2A}}$$
$$\rm{q_0\;=\;-\frac{1}{(\phi_{*}^0)^2}\;[\;2\; (1 + 2 \kappa)\; E(p)-
(1 - 2 \kappa)\;E_{2A}{p^A}\;]}$$ and the entropy reads as
\bea\label{nicee}\rm{S\;=\;\frac{2}{\phi_{*}^0}\;[\; (1 -3
\kappa)\; E(p) + (1 - \kappa)\;E_{2A}{p^A}\;]}\;\;\;.\eea Now
$\rm{\{g,g_A,g_{AB},T_{1},T_2,T_3\}}$ is given by
$$\rm{g_{AB}\;=\;-6 \;\phi_{*}^0\; \frac{1 + x- \kappa}{(1+x)^2}\;E_{AB}}$$
$$\rm{g\;=\;}\rm{\frac{x^2}{(1+x)^2}\frac{1}{\phi_*^0}\;[\;-2\;E(p)\;\{1 + x+3 \kappa + 2 \kappa
x\}\;+\;E_{2A}p^A\;\{1 + x - 3 \kappa - 2 \kappa x\}\;]}$$
\bea\label{nice4}\rm{T_{1}\;\ddot{=}\;\frac{x^2}{1+x}\;[\;\{\;1 -
\kappa \frac{3+4x}{1+x}\;\}+
(1-\kappa)\;E_{2A}p^A\;]\frac{1}{\phi_0^*}\;}\eea
$$\rm{T_2\;\ddot{=}\;\kappa\;\frac{x}{1+x}\;\frac{h}{2}\;}$$
$$\rm{T_3\;\ddot{=}\;\kappa \;\frac{x^3}{(1+x)^2}}\;\frac{E(p)}{\phi_{*}^0}\;\;\;.$$
Thus the final expression for $\rm{\Omega}$ is obtained as
$$\rm{\Omega\;\ddot{=}\;e^S\;
(\phi_{*}^0)^{\frac{h}{2}+1}\;\frac{1}{\sqrt{\det[E_{AB}]}}\;\int
dx \; (1+x)^{\frac{h}{2}}\; e^{T_{1}}
\;[\;1+\;\kappa\;\frac{h}{2(1+x)} + T_{2} + T_{3}\;]}$$
\bea\label{nice5}\rm{\;\;\;\;\;\;\;\;\;\;\;\;\;\;\;\ddot{=}\;e^S\;
(\phi_{*}^0)^{\frac{h}{2}+1}\;\frac{1}{\sqrt{\det[E_{AB}]}}\;\int
dx \; (1+x)^{\frac{h}{2}}\; e^{T_{1}}
\;[\;1+\;\kappa\;\{\;\frac{h}{2}+\frac{x^2}{(1+x)^2}\;\frac{E(p)}{\phi_0^*}\;\}\;]}
\;\;\;.\eea Now from (\ref{nicee}) and (\ref{nice4})
 it is obvious that (\ref{nice5}) does not admit the form
 (\ref{form}). Indeed if it was the case, it would be so for any $\rm{h}$, implying that
$$\rm{\int dx \; (1+x)^{\frac{h}{2}}\; e^{T_{1}}}$$ be itself of the form
 (\ref{form}), which is not the case! \footnote{The saddle-point is ${\rm x=0}$.
 At
 the leading order, the saddle approximation of (\ref{nice5}) takes
 the form of (\ref{form}), but this fails in the subleading orders.}

 As we learn from the above examples, for black holes with non-vanishing $\rm{p^0}$
 there is no guarantee that the result of (\ref{inversel}) be of the form (\ref{form}),
 if one incorporates the subleading terms of the saddle-point asymptotic
 expansion. As a consequence, even a generalization of type (\ref{inversegl})
 or (\ref{rela}) does not restore the electric-magnetic duality. That is so because
 if the corresponding metric/Jacobian-determinant does not change
 the saddle-point of the integrand in (\ref{inversel}), then it can not remove more
 than one non-invariant factor from the OSV result and if it changes the saddle-point
 of the integrand, then the leading term of the asymptotic expansion does not match
 microcanonically with the Bekenstein-Hawking-Wald entropy. It
 seems that this observation opens the possibility of deeper modifications.
\Section{An effective approach}
 The ensemble defined through (\ref{new0})-(\ref{new2}) is
canonic in as many variables as the OSV ensemble, that is by
construction the set of electric potentials is mapped one to one
to the set of new variables $\psi^\Lambda$. However since the
multiplet $(CX^\Lambda,CF_\Lambda)$ defines a vector under the
symplectic transformations it is more natural for a black hole
ensemble to treat both the magnetic and the electric charges
\emph{at the same footing}, if it is requested to produce
symplectic invariant results. In that direction, the simplest
generalization of (\ref{inversel}) is an inverse Laplace
transformation from: $Z_{invari}\equiv e^{\rm{G}}$, which
integrates over both the $\Re X \equiv \mu$ and $\Im X \equiv
\phi$ with an appropriate $\emph{Jacobian}$/measure, such that
 \bea\label{ap1}\rm{[d\mu_\Lambda][d\phi^\Lambda]\;\rm{J}(\mu,\phi)}
 \;\;\eea
defines a symplectic invariant measure. In (\ref{ap1}),
$\rm{J(\mu,\phi)}$ appears either as a Jacobian when we change
variables from those which originally define the ensemble to
$\rm{(\mu,\phi)}$ or as an intrinsic measure. As it is well known,
one choice for the measure of (\ref{ap1}) which is invariant under
the symplectic transformations is, \bea\label{measure1}\rm{
[d\mu_\Lambda][d\phi^\Lambda]\;\det(\Im F_{\Lambda\Gamma}) }\eea
where \bea\rm{F_{\Lambda\Gamma}\;\equiv\;\frac{\partial^2
F}{\partial X^\Lambda
\partial X^\Gamma}}\eea
(\ref{measure1}) is used, for example, as the intrinsic measure of
the ensemble introduced in \cite{Dewit:2005lec}. However this
measure vanishes for the case of $\frac{1}{2}$BPS black holes, so
is not universally applicable. In fact a satisfactory
\emph{universal} measure has not been presented so far. Moreover
even if we apply such a measure, to respect the electric-magnetic
duality we need to introduce a symplectic invariant free energy
$\rm{G}$. Thus regarding the fact that the OSV free energy, which
as given by (\ref{OSV}) and (\ref{FE}) forms the essence of the
the OSV proposal, is only symplectic invariant at the attractor
point, we take a more conservative approach in what follows. That
is, to enjoy the proposed relationship between the
topological-string free energy and the black hole physics within a
manifest symmetric approach, we keep the OSV free energy unchanged
but introduce an enlarged ensemble which is \emph{twice} the OSV
ensemble big, in the phase space terminology. This is done by
associating to each doublet $(\mu^\Lambda,\phi^\Lambda)$, two
canonical variables $(\chi_\Lambda,\zeta^\Lambda)$ which preserve
the Legendre transformation from the OSV free energy to the black
hole entropy. Obviously the simplest choice of such variables is
to take them linear in $\phi^\Lambda$. Thus we define an
invertible change of variables as,
 \bea\label{ap2}
 \rm{\chi_\Lambda\;\equiv\;\rm{A}(p,q)\; \phi^\Lambda\;f_\Lambda(\mu^\Lambda)\;\;\;;\;\;\;
 \xi^\Lambda\;\equiv\;A(p,q)\;\phi^\Lambda\;g^\Lambda(\mu^\Lambda)\;\;\;;\;\;\;
  \forall\Lambda}\;\;\;
 \eea
with \bea\rm{
\tilde{\ff}(\chi_\Lambda,\xi^\Lambda)\;=\;\ff(\mu^\Lambda,\phi^\Lambda)}
\;\;\; \eea
 so that the Bekenstein-Hawking-Wald entropy is given by,
 \bea\label{ap3}
  \rm{S(p^\Lambda,q_\Lambda)\;=\;\tilde{S}(\alpha^\Lambda,\beta_\Lambda)\;=\;
 \tilde{\ff}(\chi_\Lambda,\xi^\Lambda) + \pi \alpha^\Lambda
 \chi_\Lambda + \pi \beta_\Lambda \xi^\Lambda} \eea
 \bea\label{at} \rm{- \alpha^\Lambda=\frac{\partial \tilde{\ff}}{\partial
\chi_\Lambda}\;\;\;;\;\;\;- \beta_\Lambda=\frac{\partial
\tilde{\ff}}{\partial \xi^\Lambda}}\;\;\; \eea where the exact
dictionary for translating the expressions in terms of the new
charges $(\alpha,\beta)$ to those in terms of the black hole
electric magnetic charges will be given a bit later in this
section.\\
Based on (\ref{ap3}) and (\ref{at}), a black hole ensemble is
defined as, \bea\label{ap4}\rm{ \tilde{Z} \equiv e^
{\tilde{\ff}(\chi_\Lambda\;,\;\xi^\Lambda)} =
\sum_{\alpha^\Lambda,\beta_\Lambda}
\tilde{\Omega}(\alpha^\Lambda,\beta_\Lambda)\;
e^{-\pi\alpha^\Lambda \chi_\Lambda-
 \pi\beta_\Lambda \xi^\Lambda}}\;\;\;.
\eea
 As before we use the Euclidean measure for the ensemble
 (\ref{ap4}), so that
 \bea\label{ap5}\rm{d(p^\Lambda,q_\Lambda)\;\doteq\;
\tilde{\Omega}(\alpha^\Lambda,\beta_\Lambda)\;\;\doteq\;\;\int
[d\chi_\Lambda][d\xi^\Lambda] \;\; e^{\tilde{\ff}\;+\;\pi
\alpha^\Lambda \chi_\Lambda\;+\;\pi\beta_\Lambda \xi^\Lambda}}
\eea Although the measure of (\ref{ap5}),
\bea\rm{\prod_{\Lambda}d\chi_\Lambda d\xi^\Lambda\;=\;A^2(p,q)
\prod_{\Lambda} \phi^\Lambda(f_\Lambda
g^{\prime\Lambda}-f^{\prime}_{\Lambda}g^\Lambda)\;d\mu^\Lambda
d\phi^\Lambda}\eea is not \emph{universally} symplectic invariant,
unlike (\ref{measure1}), we can follow the idea of the section 4:
we require the \emph{asymptotic symplectic invariance} of
(\ref{ap5}). More precisely, given an \emph{arbitrary} order in
the saddle-point asymptotic expansion of (\ref{ap5}), we choose
the function $\rm{A(p,q)}$, in the definition (\ref{ap2}), such
that the unwanted prefactor of the corresponding OSV result is
removed.\\ The integral (\ref{ap5}) in terms of $\rm{(\mu,\phi)}$
takes the form \bea\label{ap6} \rm{d(p,q)\doteq A^2(p,q)\int
\prod_{\Lambda}[d\phi^\Lambda\phi^\Lambda]\int
\prod_\Lambda[d\mu^\Lambda(f_\Lambda
g^{\prime\Lambda}-f^{\prime}_{\Lambda}g^\Lambda)]\;Z(\mu,\phi)\;e^{\pi
A\phi^\Lambda(\alpha^\Lambda f_\Lambda+\beta_\Lambda
g^\Lambda)}}\;\;\;\eea where $\rm{Z(\mu,\phi)=e^{\ff(\mu,\phi)}}$.
Now to bridge between (\ref{ap6}) and (\ref{inversel}), we should
\emph{effectively} integrate over $\mu^\Lambda$ which fixes the
value of $\mu^\Lambda$ at $\mu_{*}^\Lambda$. Here one important
\emph{constraint} on the functions $\rm{f_\Lambda}$ and
$\rm{g^\Lambda}$ comes into play. First of all in accordance with
(\ref{at1}) we require that the saddle-point values of
$\mu^\Lambda$'s coincide with the corresponding black hole
magnetic charges,
\bea\label{match1}\rm{\mu_{*}^{\Lambda}\;=\;p^\Lambda}\;\;\;;\forall
\Lambda\;\;\;.\eea Regarding the saddle-point equation for
(\ref{ap6}), (\ref{match1}) implies that
\bea\label{match2}\rm{\alpha^\Lambda f^{\prime}_{\Lambda}(p) +
\beta_\Lambda g^{\prime\Lambda}}(p)\;=\;0\;\;\;;\;\;\;\forall
\Lambda\;\;\;.\eea Next we require that the saddle-point
evaluation of (\ref{ap6}) takes the form of \bea\label{match3}\rm{
\Omega(p^\Lambda,q^\Lambda)\;\;\doteq\;\;\int [d\phi^\Lambda]
\;\;M(p,q,\phi)\; e^{\ff\;+\;\pi\;q_\Lambda\phi^\Lambda}} \eea
which together with (\ref{match1}) imply that
 \bea\label{match4}\rm{\frac{q_\Lambda}{A(p,q)}\;=\;\alpha^\Lambda f_{\Lambda}(p)
+ \beta_\Lambda g^{\Lambda}(p)}\;\;\;;\forall \Lambda\;\;\;.\eea
The dictionary between $\rm{(\alpha,\beta)}$ and $\rm{(p,q)}$ is
now given by the solution to the equations (\ref{match2}) and
(\ref{match4}), which reads as
\bea\label{match5}\rm{\alpha^\Lambda=\frac{q_\Lambda}{A(p,q)}
\;\frac{g^{\prime\Lambda}(p)}{f_\Lambda(p)
g^{\prime\Lambda}(p)-f^{\prime}_{\Lambda}(p)g^\Lambda(p)}}\eea
\bea\label{match5prime}\rm{ \;\;
\beta_\Lambda=\frac{q_\Lambda}{A(p,q)}
\;\frac{-f^{\prime\Lambda}(p)}{f_\Lambda(p)
g^{\prime\Lambda}(p)-f^{\prime}_{\Lambda}(p)g^\Lambda(p)}\;\;\;}\eea
Now as a \emph{constraint} on the functions $\rm{f}$ and $\rm{g}$,
we require that the equations (\ref{match5}) and
(\ref{match5prime}) are consistent with the attractor equations of
(\ref{at}), for $\rm{A(p,q)}$ treated as an \emph{arbitrary given}
function. That is, the following two equations should hold
\bea\rm{\frac{\partial \tilde{\ff}}{\partial
\chi^{*}_{\Lambda}}\;=\;- \frac{q_\Lambda}{A(p,q)}
\;\frac{g^{\prime\Lambda}(p)}{f_\Lambda(p)
g^{\prime\Lambda}(p)-f^{\prime}_{\Lambda}(p)g^\Lambda(p)}}\eea
\bea\rm{\frac{\partial \tilde{\ff}}{\partial
\xi_{*}^{\Lambda}}\;=\; \frac{q_\Lambda}{A(p,q)}
\;\frac{f^{\prime\Lambda}(p)}{f_\Lambda(p)
g^{\prime\Lambda}(p)-f^{\prime}_{\Lambda}(p)g^\Lambda(p)}}\eea
with $\rm{\chi^{*}_{\Lambda}=A \phi_{*}^\Lambda
f_\Lambda(p^\Lambda)}$, $\rm{\xi_{*}^{\Lambda}=A \phi_{*}^\Lambda
g^\Lambda(p^\Lambda)}$ and
$\rm{q_\Lambda=-\frac{1}{\pi}\frac{\partial \ff}{\partial
\phi_{*}^\Lambda}}\;.$ \\
Given these requirements, the first order saddle-point evaluation
of (\ref{ap6}) takes the form of (\ref{match3}) with
\bea\label{jacob}\rm{M(p,q,\phi)\;=\;A^{2h+2}(p,q)\prod_\Lambda\;[\;
(\frac{\phi^{\Lambda}}{q_\Lambda})^\frac{1}{2}
\;\frac{(f_{\Lambda}(p)g^{\prime\Lambda}(p)-g^{\Lambda}(p)f^{\prime}_{\Lambda}(p))^\frac{3}{2}}
{(f^{''}_{\Lambda}(p)g^{\prime\Lambda}(p)-g^{''\Lambda}(p)f^{\prime}_{\Lambda}(p))^\frac{1}{2}}
\;]}\;\;\;.\eea So, the \emph{effective} inverse Laplace
transformation, (\ref{match3}), differs from the original OSV
formula, (\ref{inversel}), by a measure $M$ which in \emph{form}
is something between the metric measure of (\ref{inversegl}) and
the Jacobian-matrix in (\ref{rela}).\\ Now given a specific
prepotential and an $arbitrary$ order of the saddle-point
asymptotic expansion of (\ref{match3}), the measure (\ref{jacob})
equals
\bea\label{jacoba}\rm{\grave{M}(p,q)\;\equiv\;M(p,q,\phi_*)\;=\;A^{2h+2}(p,q)\prod_\Lambda\;[\;
\frac{(f_{\Lambda}(p)g^{\prime\Lambda}(p)-g^{\Lambda}(p)f^{\prime}_{\Lambda}(p))^\frac{3}{2}}
{(f^{''}_{\Lambda}(p)g^{\prime\Lambda}(p)-g^{''\Lambda}(p)f^{\prime}_{\Lambda}(p))^\frac{1}{2}}
\;\sqrt{\frac{\phi_{*}^{\Lambda}}{q_\Lambda}}\;]}\eea\bea\;\;\;
\rm{\tilde\Omega_{*}=\grave{M}(p,q)\;\Omega_{*}\;\;\;.}\eea  It is
obvious that $\rm{\grave{M}(p,q)}$ plays the same role as
$\rm{\det[V]}$ of the section 4 and protects the resulting black
hole degeneracy of states against the electric-magnetic duality
violations by removing the unwanted prefactors, with a proper
choice of the function $\rm{A(p,q)}$.


\Section{Conclusion}

We studied the issue of dualities in the OSV proposal. We showed
that, as far as the duality-violation of the OSV result for the
degeneracy of states is restricted to a prefactor, through a
redefinition of the ensemble in terms of proper variables, one
obtains the desired duality properties. In these cases, the
degeneracy of states which comes from the inverse Laplace
transformation with a flat measure of integration, is free from
any unwanted prefactor which appears in ordinary OSV mixed
ensemble. The result shows harmony with the microscopic results if
we are careful about the validity of the perturbative regime and
the application domain of the conjecture, at least to the extent
that we know about the microscopic results. For the case of
vanishing $D_6$-brane charge, where the ensemble is pure canonical
and the restoration of the duality is exact, our procedure shows
that the measure in the phase space of the black hole is
intrinsically flat. However this is not always the case for the
case $\rm{p^0\neq0}$. Finally from our observations the following
questions call for addressing.\\
$1.$ In all these proposals, we have no natural choice for these
variables, however, we can choose them phenomenologically.
Understanding and interpreting these variables need further
investigations. \\
$2.$ To define the ensemble of sections 3 and 4, one can absorb
the unwanted OSV prefactor within the redefinition of a
\emph{single} $\rm{X}$ variable. Among the $\rm{X}$ variables,
$\rm{X^0}$ which controls the topological-string coupling via
(\ref{topcpl}), plays a distinguished role in the prepotential
(\ref{nice0}). It is physically natural to think about a
redefinition of the OSV ensemble as $\rm{\hat{X}^0=\grave{M}(p,q)X^0}$,
according to (\ref{pos}), and ask if for a given black hole charge
multiplet $\rm{(p^\Lambda,q_\Lambda)}$, the requirement of the
electric-magnetic duality gives an effective sense to the
topological-string coupling as \emph{seen by the black hole
ensemble} through the identification $\rm{\hat{g}\;\equiv\;\frac{4
\pi
i}{\hat{X}^0}}$. \\
$3.$ Considering black holes with non-vanishing $D_6$-brane
charge, there are cases where no measure refinement as
(\ref{inversegl}) can restore the electric-magnetic duality if one
incorporates the subleading terms of the asymptotic expansion. Do
we need to go for a deeper modification of the OSV proposal?

\vspace{3cm}
\begin{center}
{\large {\bf Acknowledgments}}
\end{center}
We wish to thank Ashoke Sen for motivating us to address this
problem and for valuable discussions. We would like to thank
Cumrun Vafa and Atish Dabholkar for useful comments. A. T. thanks
Hirosi Ooguri for nice discussions during the ICTP Spring School
on Superstring Theory and Related Topics, 2005.


\end{document}